\documentclass[12pt]{article}
\usepackage{newtxtext,newtxmath}

\usepackage{graphicx}

\usepackage[letterpaper,margin=1in]{geometry}

\usepackage{lineno}

\renewenvironment{abstract}
	{\quotation}
	{\endquotation}

\date{}

\makeatletter
\renewcommand{\fnum@figure}{\textbf{Figure \thefigure}}
\renewcommand{\fnum@table}{\textbf{Table \thetable}}
\makeatother

\usepackage{scicite}

\usepackage{url}

\newcommand{\ket}[1]{\left| #1 \right\rangle}

\def\scititle{
	Unconditional full vector magnetometry using spin selectivity in Nitrogen Vacancy centers in diamond
}
\title{\bfseries \boldmath \scititle}

\author{
	Asier Mongelos Martinez$^{*,1,2}$, Jason Tarunesh Francis$^{3}$,\\
	Julia Bertero-DiTella$^{2}$, Geza Giedke$^{3,4}$, Gabriel Molina-Terriza$^{2,3,4}$,\\
	Ruben Pellicer-Guridi$^{\dagger,2}$,\\
	\small $^{1}$EHU/UPV, University of the Basque Country, \\
	\small Faculty of Science and Technology department of Electricity and Electronics, \\
	\small $^{2}$Centro de Fisica de Materiales (CFM), CSIC-UPV/EHU, Paseo Manuel de Lardizabal 5,\\
	\small 20018 Donostia-San Sebastian, Spain, \\
	\small $^{3}$Donostia International Physics Center, Paseo Manuel de Lardizabal 4, 20018 Donostia-San Sebastian, Spain, \\
	\small $^{4}$IKERBASQUE, Basque Foundation for Science, Maria Diaz de Haro 3, 48013 Bilbao, Spain\\
	\small $^*$ Corresponding author: \url{amongelos003@ikasle.ehu.eus} \\
	\small $^\dagger$ Corresponding author: \url{ruben.pellicer@ehu.eus}
}

\begin{document}
	\maketitle
\begin{abstract} \bfseries \boldmath
\noindent Vector magnetometry with nitrogen vacancy (NV) centers in diamond offers high sensitivity and room temperature operation, yet full reconstruction of an unknown magnetic field vector has remained constrained by the need for bias fields, auxiliary sensors, or polarized pumping schemes, each introducing hardware complexity, noise sources, or incomplete directional resolution. Microwave (MW) control field inhomogeneities have been exploited to overcome some of these limitations, trading the full capabilities of NV ensembles for resolving the vector magnetic field. In this work we show that homogeneous elliptically polarized MW fields alone can be exploited to individually select each of the four different NV axis directions in diamond. We present a method that relies on the knowledge of the spatial arrangement between the diamond and the microwave antenna to deterministically tailor the control fields to discriminate among the different NV axes while maintaining the benefits of NV ensemble measurements. This novel approach overcomes the inherent limitation, present in NV magnetometry, of mixing the NV orientations, without trading it for a loss in sensitivity or increased hardware complexity. We prove that our method allows to estimate  both magnitude and direction of external magnetic fields without bias fields, auxiliary sensors, or prior assumptions on the field, eliminating core constraints that have limited NV based vector magnetometry.

\end{abstract}

\noindent

Magnetic field sensors are widely used in applications ranging from communications and navigation to medical devices such as magnetoencephalography, nuclear magnetic resonance (NMR) or magnetic marker tracking \cite{lovchinsky2016nuclear, li2011quantum, vrba2002squid, yu2024noninvasive}. The quest for finding high-end, yet robust and practical, magnetometers got a new impulse using quantum control of Nitrogen Vacancy (NV) centers in diamond \cite{hahl2022magnetic}. Quantum-enhanced sensors based on NV centers take advantage of their extremely sensitive electronic spin to detect external signals. Among the observables measured using NV centers are temperature \cite{hui2019all}, strain \cite{macquarrie2013mechanical, sharma2018imaging}, electric field \cite{dolde2011electric, dolde2014nanoscale, michl2019robust} and, most prominently, magnetic field \cite{nanoscale_balasubramanian_2008,acosta2009diamonds,high_steinert_2010, hong2013nanoscale,mamin2014multipulse, magnetometry_rondin_2014, clevenson2018robust,fu2020sensitive}. The magnetic field dependent energy level structure of the ground state, dominated by the Zeeman-splitting, makes NV centers one of the most sensitive sensors for magnetic fields at room temperature \cite{barry2024sensitive, barry2020sensitivity}.

NV magnetometry relies on Electron Spin Resonance measurements (ESR) of the ground-state Zeeman splitting. A typical experimental configuration is depicted in Fig. (1 a). In bulk diamond, millions of NV centers, each oriented along one of the four crystallographic axes, enable full vector-field reconstruction (inset Fig. (1 a)) \cite{yu2025experimental}. In usual experiments, the ESR alone is unable to fully distinguish the four NV directions, as will be explained in detail below. As an example, in Fig. 1 b), (upper lines) we present two numerically calculated ESR spectra for two different magnetic fields. Existing vector-magnetometry methods solve this problem by requiring additional sensors or a bias field. This imposes constraints on the hardware, limits the dynamic range, and can perturb the sample to be measured \cite{barrera2024magnetic}. 

We introduce a protocol that uses MW fields with controlled pre-determined elliptical polarization to selectively address NV spin states. We report the formulas necessary to calculate the MW control fields that need to be applied based solely on the sensor geometry. This enables unconditional vector-field estimation—without a bias field or prior knowledge—after determination of the necessary MW field previous to using the diamond as a magnetometer. Crucially, this removes the dynamic-range limitations imposed by the need to maintain well-separated ESR peaks, significantly extending the usable field range of NV magnetometers \cite{dubey2026magnetic}.

\begin{figure}
	\centering
	\includegraphics[width=0.9 \textwidth]{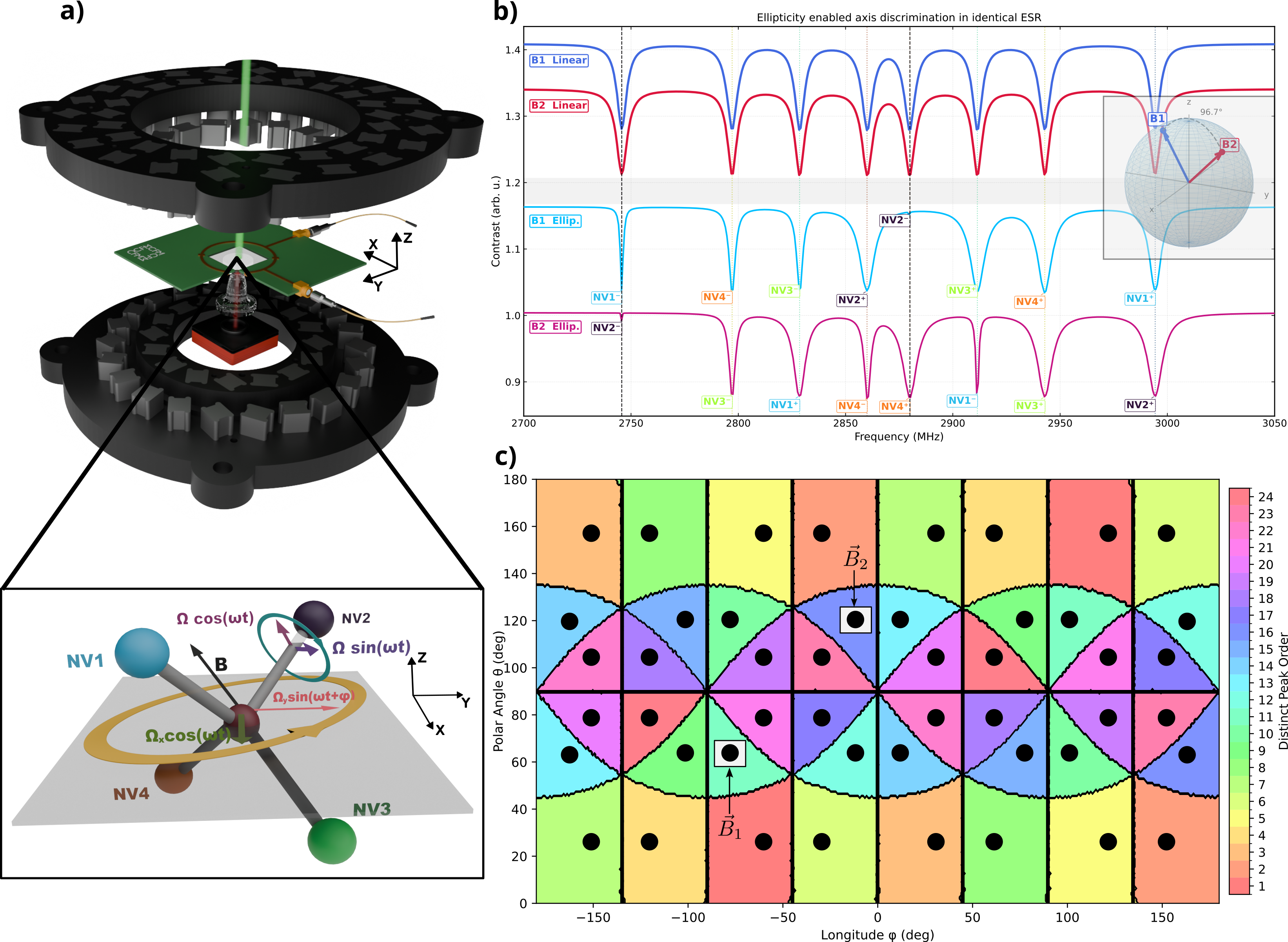}
	\caption{\textbf{Schematic depiction of spin selectivity and magnetic field degeneracy in bias-free measurements.} $\boldsymbol{a)}$ Experimental setup showing the quadrature antenna with a diamond placed in the center and surrounded by Halbach arrays of permanent magnets. Laser excitation ($532$ nm) and fluorescence collection with a parabolic compound lens and analog detector are shown. The inset shows a schematic depiction of the spin-selective unconditional magnetometry method where an elliptically polarized MW field targets one (NV2) of the four possible orientations of the NV. $\boldsymbol{b)}$  Numerical simulation of ESRs with linear (upper curves) and elliptically polarized (lower curves) MWs. Two different magnetic fields are used (B1, B2, same magnitude, different orientation as shown in the inset and in $\boldsymbol{c)}$). The two linear ESRs are identical, but the elliptical field produces different resonance patterns for the different magnetic fields, suppressing the NV2 ($m_s=-1$) line. For the linear ESR pattern there are 48 compatible magnetic fields, shown with black dots in $\boldsymbol{c)}$, which is a map of the different magnetic field directions (in polar angle coordinates). Colored areas represent zones within which the orderings of the directions and orientations producing the peaks in an ESR spectrum is preserved. The lines separating the zones indicate when the minima of two different peaks cross in the spectrum. A typical magnetometer would add a bias field so that the range of external magnetic fields to be measured remain within one zone. Here, we propose the spin selectivity shown in $\boldsymbol{b)}$ in order to lift the degeneracy }
	\label{fig:fig1}
\end{figure}
 
\subsection*{Theoretical background}
NV magnetometry protocols probe the Zeeman effect, where an external magnetic field splits the energy of the ground state spin doublet. The large zero-field splitting ($D = 2.87$ GHz) suppresses magnetic interactions in the direction orthogonal to the NV axis, up to second order perturbations \cite{barry2020sensitivity}, making NV centers highly sensitive magnetic field sensors in the direction of the NV axis. When interacting with a microwave field, the population of the spin states of the single NV orientation  ($\ket{0}$,$\ket{+1}$,$\ket{-1}$) (Fig. 1a)) can be modeled with a three-level Hamiltonian of the following form:

\begin{eqnarray} \label{GroundStateHamiltonian}
    H &=& D \, S_z^2 + \gamma_e \, \vec{B} \cdot \vec{S} + V_{MW}(t), \\
    V_{mw} &=& \Omega_x \, S_x \cos(\omega_{mw} t) + \Omega_y \, S_y \sin(\omega_{mw} t + \varphi), \label{MWInteraction}
\end{eqnarray}

\noindent where $\gamma_e$ is the electron gyromagnetic factor, $\vec{S}=(S_x,S_y,S_z)$ is the spin-1 operator vector, and $D$ is the zero-field splitting (typically $D = 2.87$ GHz). In the general form of the microwave interaction in Eq. (\ref{MWInteraction}) $\Omega_{x,y}$ are the Rabi frequencies associated with the MW field in the $X$, $Y$ quadratures respectively and $\varphi$ is the phase difference between the two quadratures. When $\omega_{mw}$ is resonant with a transition line, the effect of the MW field is to transfer the population from the zero spin state to the $\pm 1$ states. In this way, the Zeeman splitting can be observed directly from the frequency difference between the two peaks: $B_Z = \frac{f_1-f_{-1}}{2 \gamma_e}$, where $Z$ is along the NV axis and $f_{\pm 1} = D \pm \gamma_e B_Z$.

To obtain the complete magnetic vector field, i.e., the projections along $X$ and $Y$, we can use the other orientations of the NV centers. In a single bulk diamond crystal, NV centers lie in the four possible crystal directions defined by the $C_{3v}$ symmetry of the NV center point defect. In particular, in this work we have used a diamond with $[1,0,0]$ orientation in which the four possible NV axes can be described using the following set of unit vectors (as seen in Fig. (\ref{fig:fig1}a) ):
\begin{equation} \label{NV axis directions}
       \begin{aligned}
        \hat{u}_1 &= \frac{1}{\sqrt{3}}\left( 1, 1, 1\right) \ , \\
        \hat{u}_2 &= \frac{1}{\sqrt{3}} \left( -1, -1, 1\right) \ , \\
        \hat{u}_3 &= \frac{1}{\sqrt{3}} \left(  -1, 1, 1\right) \ , \\
        \hat{u}_4 &= \frac{1}{\sqrt{3}} \left(  1, - 1, 1\right) \ .
    \end{aligned}
\end{equation}

As a consequence, the ESR signal of a bulk diamond crystal in a magnetic field may exhibit up to eight distinct peaks corresponding to the four pairs of transition between states $\ket{0} \longrightarrow \ket{\pm1}$ along each of the four possible directions given by Eq. (\ref{NV axis directions}). This powerful approach allows one to perform vector magnetometry in NV centers as demonstrated by several studies \cite{sun2025omnidirectional, childress2025bias, lonard2025limits, schloss2018simultaneous}. However, the unequivocal determination of a magnetic field vector depends on the association of each peak with its corresponding NV axis \cite{clevenson2018robust, zhao2019high, wang2025fully}. Without any previous information on the magnetic field, this becomes impossible since two different magnetic fields, such as the B1 and B2 of the inset of Fig. (\ref{fig:fig1} b) can produce the same ESR spectra (red and blue lines) in a regular measurement. Therefore, a standard ESR spectrum is ill conditioned and does not contain the information needed to match each peak to its corresponding NV axis.   

In fact, due to the $C_{3v}$ symmetry of the crystal, and the existence of four different orientations, each ESR signal with all eight peaks discernible is compatible with $2 \cdot 4! = 48$ possible magnetic field solutions. This number arises from all the possible permutations of the 4 orientations of the NV centers in the ordering of the resonances in the ESR spectra, times the indistinguishability of the positive and negative spin resonances. In Fig. \ref{fig:fig1} c), all the possible magnetic field directions for a given magnetic field magnitude are presented with their azimuthal and polar angles. The configurations where the peak ordering is preserved are shown as colored areas. The example in Fig. (\ref{fig:fig1} b) is a particular case of two identical ESR spectra  (the two upper lines, blue and red), with compatible magnetic field solutions shown in Fig. (\ref{fig:fig1} c) with dots, with the actual driving fields B1 and B2 marked. Traditionally, this problem is solved by applying a known bias field which, even in the absence of an external field, splits the eight peaks in a known configuration \cite{wang2025fully, clevenson2018robust, barry2020sensitivity}. Then, much weaker external magnetic fields would shift the position of the peaks, resulting in small variations of the ESR signal which can be measured with lock-in amplifiers enabling the complete resolution of the vector magnetic field \cite{schloss2018simultaneous, yu2025experimental, zheng2024integrated}. 

The bias field imposes a restriction on the available dynamic range of the NV centers as sensors, as the sum of the external and bias fields should always be limited to a single colored area in Fig. (\ref{fig:fig1} c).  This restricts the use of NV-center-based sensors in situations requiring measurements of magnetic fields that may vary in orientation and magnitude. Without considering the linewidth of the resonance peaks in an ESR spectrum we estimate the biggest field difference in a particular direction that can be measured under a bias field ($B_0$) is $\sim 0.5 B_0$. This would correspond to the bias field being in the centroid of one of the coloured regions in Fig. (\ref{fig:fig1} c) and having a resultant field in the border of the region furthest to it. Considering any direction of the target field, the biggest field that will not change the peak distribution is reduced to $\sim 0.2 B_0$. As shown in \cite{dubey2026magnetic} this nominal magnetic field range of measurement is reduced even further when a realistic linewidth of the resonance peaks is considered. Therefore, to measure a wide range of magnetic fields the amplitude of the bias field $B_0$ has to be increased. However, this is not possible in many interesting magnetometry applications involving magnetically sensitive samples, as the sensor itself may affect the natural state of the samples being observed. This  critically hinders measurements in studies with magnetic nanoparticles in cells or magnetic materials \cite{barrera2024magnetic, backlund2017diamond, bae2011effect, maximov2024enhanced, zakutna2020field, selim2023review}. As a consequence it becomes necessary to develop magnetometry protocols that exploit the sensitivity of NV ensembles without limiting the magnetic field range that can be measured and without interference with the samples being measured. \\

In this work we propose to use the full potential of the microwave field to deterministically address different spin states corresponding to different NV orientations, as shown in Fig. \ref{fig:fig1}. Here, it can be observed how the MW field imposed in a) specifically suppresses the lines NV2- of the spectra, as shown in Fig. \ref{fig:fig1} b) with cyan and magenta lines (two bottom lines). This allows to discriminate between the fields B1 and B2, because different resonances behave differently for an elliptically polarized MW field, acting as a labeling for the different peaks of the ESR spectrum. From Eqs. (\ref{GroundStateHamiltonian}-\ref{MWInteraction}) if $\Omega_x=\Omega_y$ and $\varphi=0$, i.e., a circularly polarized MW field is applied, the transfer of population only happens in $\ket{0} \longrightarrow \ket{+1}$, while blocking the $\ket{0} \longrightarrow \ket{-1}$ (Supplementary Text). In this work we propose to use that effect to label a particular transition with its parent axis. We entirely avoid the use of a bias field by calculating the driving fields for each crystal direction, using only the knowledge of the orientation of the crystal with respect to the antenna. This allows for the use of the full dynamic range of NV sensors and makes them competitive magnetometers. Spin selectivity using the microwave polarization and orientation has been previously proposed \cite{lenz2021magnetic, kumar2023high, kolbl2019determination, childress2025bias}, as well as using  a single-axis configuration to demonstrate axis selectivity for a particular scenario of a diamond cut in [111] \cite{herrmann2016polarization, mrozek2015circularly, alegre2007polarization, zheng2019zero, li2023near, staacke2020method}. In \cite{yoon2025identifying} the spatial variation of the MW field produced by a loop antenna is used to discriminate different axis orientations, which relies on spatial inhomogeneity of control fields which limits the benefits offered by ensemble measurements. In this work, we exploit these properties with a fully fledged protocol that makes use of analytical calculations to determine the ideal MW control fields in a wide interrogation area that exploits all four crystal directions simultaneously in a diamond cut at [100], which can be readily used for vector magnetometry.

\subsection*{Spin-selective vector magnetometry}
To understand the protocol, note that a MW field that is circularly polarized for one direction in Eq. (\ref{Circular vectors}) becomes elliptically polarized for the others. Because transition selectivity requires perfectly circular MW fields, only the NV axis aligned with the designed polarization will lose one of its transition lines, while the other axes still show two peaks (with possibly different contrast). This allows us to label each pair of ESR peaks by tailoring the MW field to be circular for a chosen axis. Moreover, the handedness of the MW polarization determines the sign of the magnetic-field projection on that axis, indicated by whether the attenuated peak appears above or below the zero-field splitting at $2.87$ GHz.

The technical challenge then is to produce the required MW fields for each of the NV orientations using antennas that are compatible with magnetometry measurements. A typical antenna that is well-suited for these kinds of instruments would be a resonant planar antenna. In our system, we have used a 2-port quadrature PCB antenna capable of generating arbitrary MW signals with great homogeneity \cite{pellicer2025versatile}. These antennas can only control the polarization in the plane of the antenna. Nevertheless, as we will show below, this is enough to control the spin selectivity in the experiments. First, one starts by defining the set of right- and left-circular vector fields ($\vec{r_i},\vec{l_i}$) which will address each of the $\hat{u}_i$,  $i=1, ..., 4$ orientations: 
\begin{equation} \label{Circular vectors}
    \begin{cases}
        \vec{r_i} = \cos(\omega t) \vec{M_i} + \sin(\omega t) \vec{N_i}, \\
        \vec{l_i} = \cos(\omega t) \vec{M_i} - \sin(\omega t) \vec{N_i},
    \end{cases}
\end{equation}
where $\Vec{M_i}, \Vec{N_i}$ define 2 orthonormal vectors to  the $i$-th axis so that $\hat{u_i} \cdot \vec{M_i} = \hat{u_i} \cdot \vec{N_i} = \vec{M_i} \cdot \vec{N_i} = 0$. If we position the antenna to lie in the $XY$-plane, we will then produce fields that may have a component along the NV axis:  
\begin{equation}
    \begin{cases}
        \vec{V_i} = \left( a \vec{r_i} + b \hat{u_i} \right), \\
        \vec{T_i} = \left( a \vec{l_i} + b \hat{u_i} \right) \ \ .
    \end{cases}
\end{equation}
where the coefficient $a$ is the strength of the microwave field generating the spin rotations, and thus, proportional to the Rabi frequencies of the NV ($\Omega_{x,y}$ in Eq. (\ref{MWInteraction}) ).  The parameter $b$ is then chosen so that the vectors $\vec{V_i}$ and $\vec{T_i}$ lie in the plane of the antenna, thereby producing an elliptically polarized MW field in the plane of the antenna but circularly polarised in the plane orthogonal to NV axis $i$ (Supplementary Text). This approach is completely independent of the external magnetic field and relies only on knowledge of the diamond cut and its position in the antenna plane. 

For example, if the diamond is rotated around the $X$ or $Y$ axes with respect to the plane of the antenna, the previous reference frame becomes tilted. To take this into account and obtain the expressions of each polarization in the laboratory reference frame the previously achieved vectors need to be rotated. This rotation is achieved applying rotation matrices $R_x(\phi)$ and $R_y(\theta)$. Interestingly, small tilts in the diamond with respect to the antenna are an effective tool to lift the degeneracy created by the symmetry of the crystal (see \cite{munzhuber2020polarization} and Supplementary Text). This allows to  deterministically calculate a set of microwave polarizations that will select exactly each of the peaks created by each axis direction. Solving for the optimal parameter (see Supplementary Text for the complete derivation), the formulas for the signal in the $X$ and $Y$ directions that produce an optimal circular pulse for NV orientation $i = 1, ..., 4$ are:

\begin{equation}\label{signal port X formula Main}
\begin{split}
    S_i^{X} &= a \frac{1}{ \sqrt{2} n^{x,y}_i (\sin (\phi )+ q_i \sin (\theta ) \cos (\phi ))+2 n^z_i \cos (\theta ) \cos (\phi )} \times \\
            & \quad \Big[ s_{i} \sqrt{2} \sin (\omega t) \left(\cos (\phi )+ p_i \sin (\theta ) \sin (\phi )\right) + \\
            & \quad \quad k_i\cos (\omega t) \left(2 n^{x,y}_i \cos (\theta ) \sin (\phi )+\sqrt{2} n^z_i (\sin (\theta ) \sin (\phi )+\cos (\phi ))\right) \Big] \ ,
\end{split}
\end{equation}
\begin{equation}\label{signal port Y formula Main}
    S^Y_i = a \frac{2  n^{x,y}_i \sin (\theta ) \cos (\omega t)+ g_i \sqrt{2}  \cos (\theta ) (n^z_i \cos (\omega t)+ h_i \sin (\omega t))}{ \sqrt{2} n^{x,y}_i (\sin (\theta ) \cos (\phi )+\sin (\phi ))+ f_i 2 n^z_i \cos (\theta ) \cos (\phi )} \ ,
\end{equation}
where $a$ is a normalization constant, $n^{x,y,z}_i$ are the components of the direction of axis $i$, $ \hat{u}_i = \left( n^x_i, n^y_i, n^z_i\right)$ and $s_i, p_i, q_i, g_i, h_i, f_i$ are coefficients that take values $\pm1$ which are listed in the Supplementary Text together with the complete derivation.

Equations (\ref{signal port X formula Main}) and (\ref{signal port Y formula Main}) show the  MW polarizations that would completely attenuate the peak corresponding to a particular axis. These general expressions show that for any diamond (cut in the $[100]$ direction) tilted by some angle with respect to the antenna it is possible to find a set of different polarizations that will generate circular polarizations around each of the axes deterministically. Each of these polarizations is directly tied to the extinction of one of the resonant peaks of one axis while maintaining the 2 resonant peaks for all the other axes. Therefore, the directional resolution of the vector external magnetic field is possible using only the knowledge of which MW field we are using in the $XY$ plane, as will be shown in depth later. \\


\begin{figure}
	\centering
	\includegraphics[width=\textwidth]{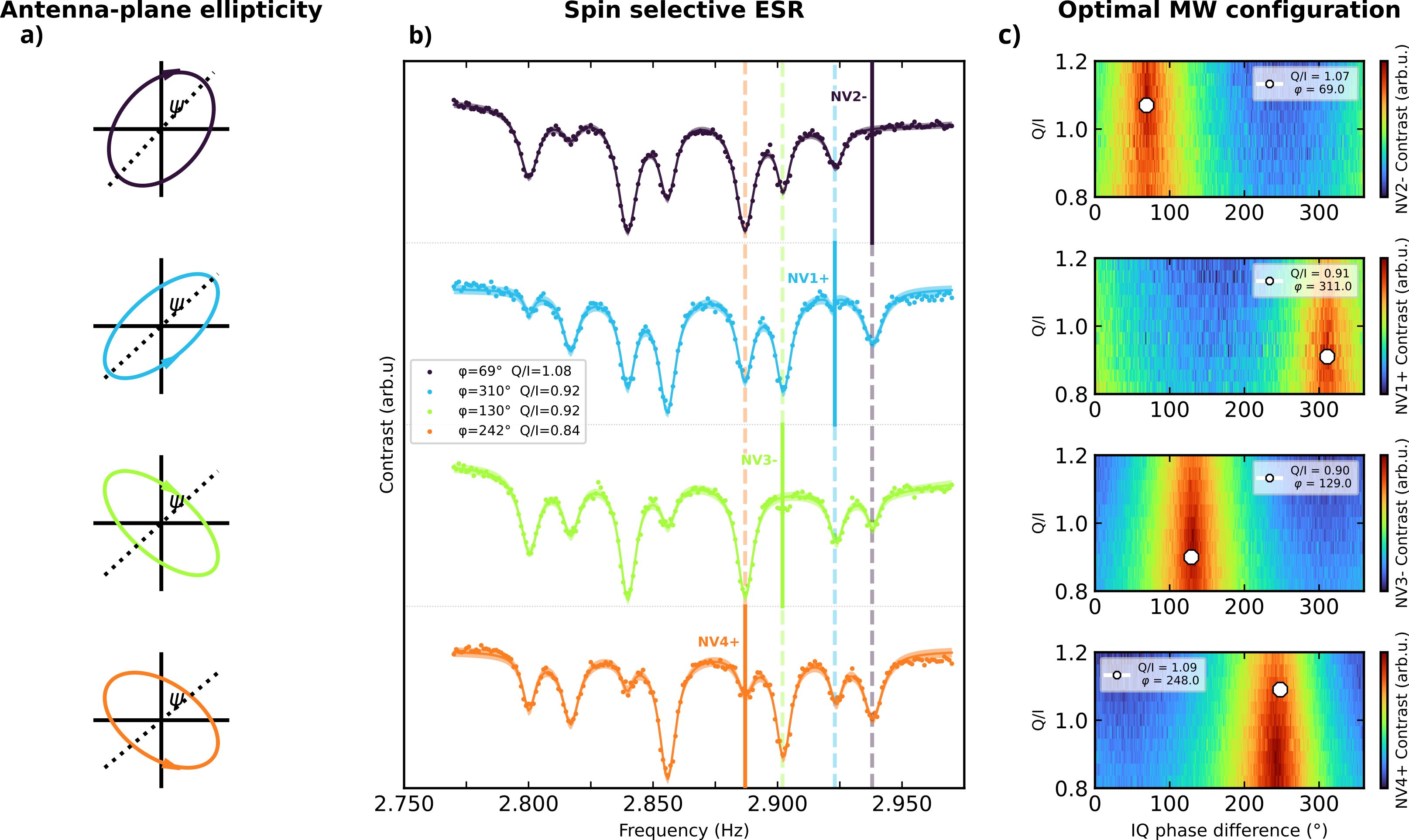}
	\caption{{\textbf{Experimental proof of concept, where each row shows the MW fields to target individual NV directions NV2 (-), NV1(+), NV3(-) and NV4(+) from top to bottom where (+/-) refer to the tranisiotn correponding to $m_s=+1$ or $-1$ respectively}. $\boldsymbol{a)}$ MW field ellipses generated by the Arbitrary Waveform Generation (AWG) optimal for each crystal direction. $\boldsymbol{b)}$ ESR spectra driven by the corresponding MW field in $\boldsymbol{a)}$, showing attenuation of a single NV axis enabling the identification of the peak distribution in the ESR spectrum. $\boldsymbol{c)}$ Full sweep of the MW fields showing the dependency of peak attenuation to MW phase and amplitude. The configurations resulting in the highest contrast correspond to the results in a) and b) which are in agreement with the calculated optimal configurations marked as white dots. The attenuation of the peaks will vary only slightly with small imperfections such as the crystal tilt, inhomogeneities of the MW field and power imbalance between the ports of the antenna. The system is more sensitive to the phase difference. Overall, this proves the deterministic nature of the method and its robustness under small imperfections.}}
	\label{fig:fig2}
\end{figure}
\subsection*{Results}

In Fig. \ref{fig:fig2} we show experimentally how the spin selectivity is achieved. As mentioned before, a tilt in the diamond crystal becomes a useful design parameter that can be engineered to optimize sensor performance. For our experimental conditions we use $\theta \sim -10^{\circ}$, $\phi \sim 1^{\circ}$ which using equations  (\ref{signal port X formula Main}) and (\ref{signal port Y formula Main}) results in the MW polarization  shown in in Fig. (\ref{fig:fig2} a). From the formulas, these polarizations are achieved for this set of phases for NV2-, NV1+, NV3-, NV4+: $69^\circ$, $311^\circ$, $129^\circ$, $248^\circ$ respectively. (See Supplementary Text for the exact phase and amplitude parameters for this experimental setup). In Fig. (\ref{fig:fig2} a) and (\ref{fig:fig2} b) we present the optimal MW polarizations and their corresponding ESR spectra. One can observe how each of the polarizations selectively addresses one pair of peaks, diminishing the visibility of one of them. Moreover, we studied the behaviour of the contrast of the peaks with a scan of the relative amplitude and the relative phase of the antenna feeds, denoted as the I and Q channels for MW in the $X$ and $Y$ directions, respectively. This allows us to better understand the effect of hardware imperfections to our deterministic approach. This is seen in Fig. (\ref{fig:fig2} c).  The deterministic optimal MW parameters derived from equations  (\ref{signal port X formula Main}) and (\ref{signal port Y formula Main}) are marked as white dots in Fig. (\ref{fig:fig2} c). The biggest difference in calculated and measured optimal phase is of $6^\circ$, corresponding to NV4+. This is in a remarkable agreement between the theoretically expected values and the experimental results. The biggest errors may be due to the uncertainty in the exact geometry of the experiment  together with experimental MW generation imperfections. 



The narrow optimal regions along the phase axis reflect the extreme sensitivity of the approach to the phase of the MW \cite{zheng2019zero, li2023near}, making it possible to identify the correct phase configuration for all axes. Remarkably, this shows the deterministic nature of the protocol: The MW fields obtained through the geometry of the setup Fig. \ref{fig:fig2} a) are directly linked to the attenuation of one of the ESR peaks (Fig. (\ref{fig:fig2} b) and (\ref{fig:fig2} c) ) which indicates the strength of the magnetic field along that axis. We will prove later that this is sufficient to perform full vector magnetometry.  



From an independent characterization of the magnetic field we expected NV2 to have the biggest field projection followed by NV1, NV3 and NV4. Moreover, the projections of axes NV4 and NV1 were expected to be positive and those of NV3 and NV2 negative. This is also confirmed experimentally by the optimal handedness of the fields needed to attenuate the peaks with frequencies above $2.87$ GHz. As shown by the ellipses in Fig. (\ref{fig:fig2} a) the peaks with frequencies higher than $D=2.87$ GHz for axis NV2 and NV3 are attenuated by a left-handed MW field. This means the peaks correspond to the $\ket{0} \longrightarrow \ket{-1}$ transition and thus the field projection in these axes is negative. Directions NV4 and NV1 use right-handed polarization owing to the positive projection of the field over them. The results are consistent with an independent field measurement and confirm the possibility of labeling NV axes with their corresponding transition peaks and extracting the sign of the projection of the field over them.

\begin{figure}
	\centering
	\includegraphics[width=0.8\textwidth]{figures/Measurement_Protocol_gmt_v8.png}
	\caption{{\textbf{Unconditional Spin-Selective Vector Magnetometry.} $\boldsymbol{a)}$ Visual depiction of the main steps of the measurement protocol, retrieval of the frequencies for four ESR measurements with pre-calibrated spin-selective MW configurations and calculation of the magnetic field. $\boldsymbol{b)}$ Flow chart of the specific protocol used in our experiment.}}
	\label{fig:fig3}
\end{figure}

Having demonstrated that spin selectivity allows to label each one of the peaks in the ESR spectrum, we now propose a protocol to produce bias-free high-dynamic-range vector magnetometry. The protocol we propose is represented in the scheme in Fig. \ref{fig:fig3}. The protocol includes a preliminary step, where a calibration of the antenna can be performed to determine that the instrument produces the elliptical fields needed in the following steps. This step is performed once, independently of any external magnetic field and requiring only the determination of the antenna's response to external currents. This calibration can be performed under laboratory conditions and can be repeated to account for environmental factors slightly changing the antenna performance. Once the right MW fields are produced to address specific orientations, this information is used to define the MW antenna feed conditions and to post-process the data. Using four MW configurations with their respective handedness provides a sufficient measurement basis. For any ESR spectrum the peak attenuated with each MW configuration corresponds to a given axis and handedness. When all axis transition lines and signs have been identified (step two in Fig. \ref{fig:fig3}) the field projection vector can be constructed. Then, knowing the axes orientations with respect to the antenna, the vector magnetic field in the laboratory frame can be calculated using techniques such as the Moore-Penrose pseudo-inverse or best linear unbiased estimator (BLUE) of the $\mathbf{A}$ matrix (step three in Fig. \ref{fig:fig3} a) ) \cite{lonard2025limits, schloss2018simultaneous}.

\begin{figure}
	\centering
	\includegraphics[width=0.8\textwidth]{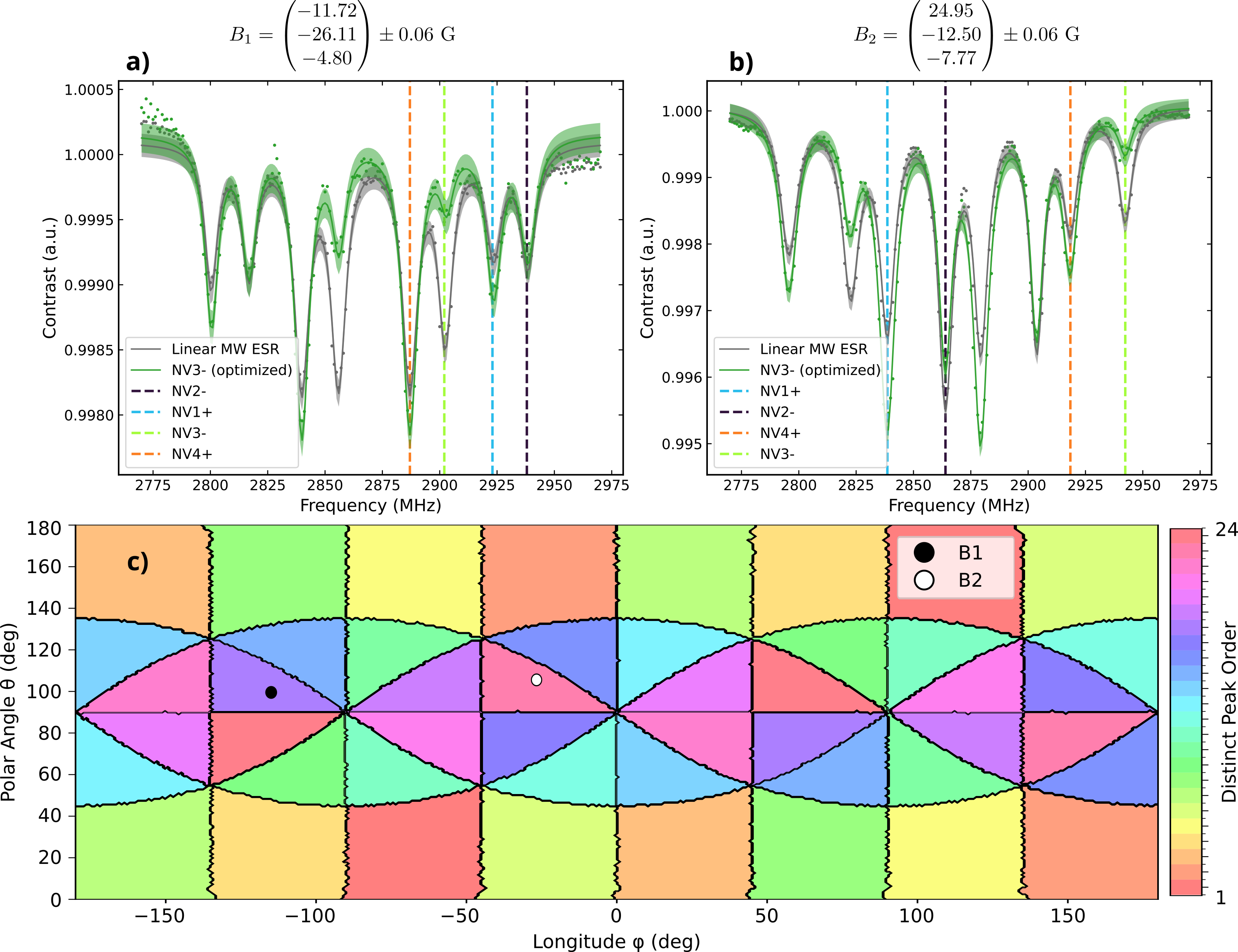} 
	\caption{{\textbf{Unique solution of magnetic fields with similar ESR signature.} $\boldsymbol{a)}$ ($\boldsymbol{b)}$) ESR measurement under magnetic field $\vec{B_1}$ ($\vec{B_2}$) with linear MW field (black line) and the optimized MW control field for NV3- (green line). $\boldsymbol{c)}$ Location of the resolved magnetic vectors in the configuration map of bulk NV centers, showing the calibration and protocol presented are sufficient to resolve magnetic fields generating arbitrary peak orientations with similar ESR signatures.}}
	\label{fig:fig4}
\end{figure}

In order to prove that our protocol can discriminate two fields which would in principle be  indistinguishable in typical NV-based magnetometry, we performed a final test. In Fig. (\ref{fig:fig4} a) and (\ref{fig:fig4} b) we show measurements for two different magnetic fields with the same magnitude, but pointing in two different directions. The two external fields  belong to two different sectors in the magnetic field configuration map, as shown in Fig. (\ref{fig:fig4} c). These two fields produce nearly identical ESR spectra under a linear MW field excitation (black lines in the ESR spectra). However, by using the spin selectivity, we can identify that the two fields have different projections onto the NV axes, achieving the discrimination of the magnetic fields and being able to measure their vectorial components.  As an example, in the figure we plot in green the ESR spectra for elliptical polarizations addressing NV3-. Using the four MW polarizations shown in Fig. \ref{fig:fig3} we obtain four different spectra associated with the projections of the magnetic field. The resultant laboratory-frame magnetic field is $\vec{B_1} = (-11.72,\ -26.11,\ -4.80) \pm 0.06 $G for Fig. (\ref{fig:fig4} a) (black dot in Fig. (\ref{fig:fig4} c)) and $\vec{B_2} = (24.95,\ -12.50,\ -7.77) \pm 0.06 $G  for the field used in Fig. \ref{fig:fig4} b) (white dot in Fig. (\ref{fig:fig4} c)). This is a proof of the validity of our vector magnetometry protocol based on spin selectivity, demonstrating the capability to lift the degeneracy of the 48 possible fields which would produce the same linear spectra.

\subsection*{Discussion}

The proposed method can unconditionally resolve the direction of the magnetic field, which is a crucial advantage over prevailing approaches. As mentioned before, the standard solution requires a bias field to determine the magnetic field. Our approach not only removes the errors introduced with hardware precision limitations of the bias, but also enables measuring the magnetic field sources in their natural state \cite{bae2011effect}. Also, one important advantage as mentioned in the theoretical description is that the dynamic range of the sensor is increased by orders of magnitude. For example, in order to cover the measurement of both magnetic fields $\vec{B_1}$ and $\vec{B_2}$ of Fig. \ref{fig:fig4} c)  with the bias field method and the same sensor, a bias field of at least $80$G should be present, so that the addition of the bias and external fields for both $\vec{B_1}$ and $\vec{B_2}$ would fall in the same colored section of Fig. (\ref{fig:fig4} c). Then, the lower limit of measureable fields would be determined by the linewidth of the ESR spectra, i.e. magnetic fields of the order of $10^{-2}$G. Better sensitivities could be achieved, but they would require more complex methods affecting the upper limit of the dynamic range \cite{clevenson2018robust}. In this case the upper limit of the dynamic range of the sensor would be around half of the bias field (in the best-case scenario that the target fields are in a particular direction with respect to the bias), as explained earlier, spanning up to external magnetic fields of $40$G approximately. 

To estimate the dynamic range of our measuring protocol, let us take a closer look to Fig. \ref{fig:fig2} b). A noteworthy effect occurs at the inner peaks (close to $2.87$ GHz) where less attenuation of the peaks is achieved. We attribute this effect to dipole coupling between NV centers and strain effects in high-NV-density diamond samples under low magnetic fields \cite{tyler2023higher, balasubramanian2008nanoscale}. For the inner peaks, the handedness of the elliptical polarization determines the attenuation of each peak, reflecting the sign of the field projection. This allows us to estimate some limits on the magnetic field magnitudes that can be resolved with our approach. The lowest magnetic fields that can be measured in this scheme is given by $\gamma_e B_z >\Pi$, where $\Pi$ is the magnitude of the transverse strain coupling. On the other hand, one can estimate the highest magnetic field that can be measured in this scheme when the transverse magnetic field is not negligible anymore, which occurs whenever $\gamma_e B_\perp \sim D$  \cite{barry2020sensitivity}. In this scenario, the approximation used for the analytical solutions of the optimal MW control fields breaks and different MW fields would need to be calculated which is out of the scope of this work. For the diamond samples used in this work the transverse strain magnitude was estimated to be $\Pi \sim 1$ MHz, which results in an estimated dynamic range of the sensor ranging from $3 \cdot 10^{-2}$ to $10^3$G, spanning 4 orders of magnitude, unconstrained in field direction. 

Our method is based on exploiting the spin selectivity in the NV using different MW polarizations. Another possibility would be to use the polarization of the excitation light. This approach exploits the coupling of the electrical NV dipoles to the excitation laser beam depending on the orientation axis to cancel the contribution from certain NV axis to the ESR spectrum. Different studies have made great advances in discrimination of NV orientation in an ESR exploiting the excitation laser polarization \cite{backlund2017diamond, munzhuber2020polarization, li2024vector,chen2020calibration,zheng2019zero}. Potentially, a similar protocol than the one used in this work, could be adapted to deal with the excitation light polarization.  However, reports exploiting optical selectivity often trade many of the benefits of using NV ensembles for the possibility of performing vector magnetometry. For example \cite{backlund2017diamond, munzhuber2020polarization, chen2020calibration} focus the excitation beam using microscope objectives, which limits the interrogation area to small ($\sim \mu \text{m}^2$ sized) areas. This limits the number of sensors being simultaneously interrogated. Additionally, tailoring excitation polarization to decouple from a certain orientation reduces the overall fluorescence intensity as one fourth of the emitters are inactive \cite{barry2024sensitive}. This also affects spin initialization times on NVs, since different orientations lead to different excitation rates, which needs to be considered for sequences that use background measurements to reduce noise and pulsed sequences in general. All this translates to a loss of sensitivity, which is one of the main appeals of magnetometry with NV ensembles \cite{barry2020sensitivity}. In our sensor, the main sensitivity loss is due to inhomogeneities on the ellipticity of the MW field generated by the antenna. One can observe in Fig. \ref{fig:fig2} b) that although a great reduction in the visibility of the selected peaks is achieved, they are not completely attenuated, as would be expected when perfectly circular polarization is generated. This is due to the large excitation volume that is being used in this experiment ($\sim 0.5 \ \text{mm}^3$). This imperfect behavior is expected due to the sharp dependence of optimal attenuation conditions to phase, and the slight inhomogeneity of ellipticity being produced in the antenna volume (see reference \cite{pellicer2025versatile} for details). Strict peak extinction in a smaller interrogation volume using a confocal microscope setup is shown in the Supplementary Text. However, here we wanted to test the method in a realistic scenario in which large diamond volumes are used to take advantage of a higher SNR \cite{barry2020sensitivity} and applicability of the method.

Another drawback of reported optical selectivity methods is that they do not discriminate between positive and negative orientations. To measure an external magnetic field orientation uniquely it is necessary to know if the projection of the field along a particular axis is positive or negative. In \cite{munzhuber2020polarization} they overcome this by combining optical polarization with MW polarization. However, they rely on MW field inhomogeneities over the sample as the source of information to construct the magnetic vector. In practical situations, this means that the manipulation of NVs over the interrogation volume would be inhomogeneous and the sensitivity will decrease with respect to the theoretically predicted limits. Additionally, this method has stricter hardware constraints as it needs to carefully combine the hardware for MW polarization with the hardware of light polarization. In this work we show how homogeneous MW control fields can be deterministically tailored to be used in large ($\sim mm^2$) interrogation areas to discriminate positive and negative resonances coming from distinct NV orientations as well as magnetic field projection signs.

In this work, we have aimed to prove the possibility of performing spin-selective vector magnetometry relying only on spin-control MW fields, without optimizing the measurement protocol for a particular use case. Regarding the application and versatility of the method, we acknowledge that several areas of improvement exist. A limitation of the presented method is that most of the antennas being used currently are linearly polarized  and these lack the versatility to generate elliptical polarization, as needed for this method.  Although, PCB-based quadrature antennas are inexpensive and readily available, tailored designs may be needed in some cases to accommodate for spatial constraints specific to certain instruments. Another factor limiting the versatility lies in the use of arbitrary waveform generator (AWG) devices, which increases the cost of  any vector magnetometer based on this protocol beyond commercial feasibility. However, since the protocol only requires four pre-defined MW phase configurations for the resolution of the magnetic vector, standard signal generators with inexpensive programmable phase shifters and attenuators could be used to generate pre-defined ellipticity configurations. The use of phase shifters instead of bulky hardware also paves the way for the use of the protocol in miniaturized devices. With respect to the current limitation of all NV magnetometry methods, including ours, of measuring near zero magnetic fields, using less doped samples with low strain would lessen the effect of the dipolar coupling between NV centers reducing the non-linear response range.

Regarding the sensitivity of our approach, typical NV magnetometers using linear polarization of the MW fields and a magnetic bias field only need one ESR spectrum. In contrast, our method requires to measure spectra for four different pre-defined elliptical configurations, which allows it to resolve the direction of the magnetic field. A similar approach was taken in Ref. \cite{childress2025bias} were the authors used the different couplings of a linearly polarized MW field to distinguish the signals coming from the four orientations in the frequency analysis of a pulsed system. Our approach can be adapted to that protocol or any other pulsed protocol, delivering  deterministic distinguishability of the four NV orientations. Importantly, although the need to take four measurements seems a priori detrimental for the sensor response, all these acquisitions add signal constructively and can be used to enhance the SNR akin to averaging. The protocol in this work has not been optimised for delivering the ultimate sensitivity, including the post-processing algorithm and SNR conditions. These processing choices can greatly affect the sensitivity of the measurements. Nonetheless, given that when one peak is attenuated, other peaks are enhanced, we expect that on average the SNR loss should be similar to other single axis protocols. In-depth analysis of optimal configurations may depend on diverse conditions such as the crystal orientation of the NV with respect to the MW antenna. Such optimization study is  an ongoing research effort of the group and is beyond the scope of the current manuscript.

\subsection*{Conclusion}

We have presented a method for unconditional spin-selective, bias-free, high-dynamic-range magnetometry with electron spin resonance of NV-center ensembles. The protocol is based only on pre-defined elliptically polarized MW fields that depend solely on the geometry dictated by the diamond crystal - MW antenna combination. This provides the means to take full advantage of the complete dynamic range provided by the Zeeman splitting of electrons in NV centers. Moreover, we measure the response of NV ensembles to MW field dephasing and show the sharp sensitivity of NV centers to phase changes of MW signals. 

Removing the need of bias fields is particularly important when the system under measurement or its environment does not have a linear response to magnetic fields. Similarly, the methods introducing a bias field rely on the magnitude of the measured field not being larger than the bias, constraining the dynamic range. On the other hand, the works that employ additional traditional magnetometers to solve for the degeneracy of their method are only valid for far-field measurements, preventing the use of highly sought applications that exploit the exquisite spatial resolution provided by NVs. Finally, works using optical polarization can not fully resolve between the $\pm1$ states and often use small interrogation volumes, while MW inhomogeneity for axis orientation selection limits the coherent control capabilities which negatively impacts the possibility of producing sensors that achieve the theoretical optimal sensitivity. Our approach relies on coherent modulation of large NV ensembles with deterministically pre-calculated homogeneous MW fields, therefore avoiding the trade-off between axis orientation information and reduced SNR/sensitivity.

The method proposed here is versatile and can be adapted to different situations. We have described its limitations in sensitivity and dynamic range for NVs and have demonstrated the method in a situation where the external magnetic field resulted in four pairs of resonances in the ESR spectrum. Our method can also be used in situations where the orientation of the magnetic field results in a more degenerate spectrum with fewer peaks, as it would allow us to determine the orientation related to the overlapping peaks and determine its position more accurately by suppressing the other resonance. We leave a detailed study of this and other situations for a future work. We also point out that the approach utilized here is applicable to any solid state spin system with a similar dynamics given by the Hamiltonian presented here. The vectorial magnetic field could be retrieved if enough orientations of the spin can be found in the crystal, and our method could be adapted to such orientations. The four orientations used here are inherent to single-crystal diamond lattice structures with implanted point-defects. 

We conclude by remarking that the novel approach described in this work maintains the absolute precision and high spatial sensitivity nature of NVs while significantly reducing hardware constraints as compared with prevailing methods enabling easy implementation of a new class of magnetometry protocols.



\clearpage 

%
\bibliography{Ellipticity_biblio} 
\bibliographystyle{sciencemag}

%
%
%
%
%
%


\section*{Acknowledgments}
\paragraph*{Funding:}

A.M.M., J.T.F., J.B., G.M.T. and R.P.G.  acknowledge the support
from CSIC Research Platform on Quantum Technologies Grant
No. PTI-001, from IKUR Strategy under the collaboration agreement between Ikerbasque Foundation and DIPC/MPC on behalf
of the Department of Education of the Basque Government, from Project No. PID2022-143268NBI00 of Ministerio de Ciencia, Innovación y Universidades, from Elkartek grant for the BetiSense project (KK-2025/00107) from the Basque Government, from Gipuzkoa Quantum (2025-QUAN-000022-01) from the Diputación de Gipuzkoa, from Proyectos de Generación de Conocimiento (PID2024-160158OA-I00) from the Spanish Agencia de Investigación Estatal and Ramón y Cajal fellowship (RYC2023-044021-I) from the Spanish Agencia de Investigación Estatal. A.M.M. also acknowledges the financial support from the  Department of Education of the Basque Government through the Non-Doctoral  Researcher Pre-Doctoral Formation Program, Grant No. PRE\_2024\_2\_0180. G.G. acknowledges the grant PID2023-146694NB-
I00 funded by MCIN/AEI/10.13039/501100011033.

\paragraph*{Author contributions:}
A.M.M., R.P.G. and G.M.T. conceived the idea of using elliptical MW fields to target individual NV directions. A.M.M., R.P.G., J.T.F. and J.B. prepared the experimental setup and performed the measurements. A.M.M. , R.P.G. and G.M.T. analyzed the data and wrote the manuscript.  A.M.M., R.P.G., G.M.T., J.T.F., J.B., G.G. discussed and interpreted the results and provided feedback for the manuscript. A.M.M. performed simulations. G.M.T. and R.P.G. provided administrative help and funding for the work.

\paragraph*{Competing interests:}
There are no competing interests to declare.

\clearpage

\renewcommand{\thefigure}{S\arabic{figure}}
\renewcommand{\thetable}{S\arabic{table}}
\renewcommand{\theequation}{S\arabic{equation}}
\renewcommand{\thepage}{S\arabic{page}}
\setcounter{figure}{0}
\setcounter{table}{0}
\setcounter{equation}{0}
\setcounter{page}{1} 


\begin{center}
\section*{Supplementary Materials for\\ \scititle}

Asier Mongelos Martinez$^{*}$, Jason Tarunesh Francis,\\
Julia Bertero di Tella, Geza Giedke, Gabriel Molina-Terriza, Ruben Pellicer-Guridi$^{\dagger}$,\\
\small $^*$ Corresponding author:\url{amongelos003@ikasle.ehu.eus} \\
\small $^\dagger$ Corresponding author: \url{ruben.pellicer@ehu.eus}
\end{center}

\subsubsection*{This PDF file includes:}
Materials and Methods\\
Sections S1 to S4\\
Figures S1 to S5\\
Table S1 and S2

\newpage


\subsection*{Materials and Methods}

To prove the concept experimentally the quadrature antenna with the diamond in the center is placed between two rings containing permanent magnets arranged in $8$ Halbach arrays that control the external magnetic field. Laser excitation ($532$ nm) from a multimode fiber impinges from the upper plane of the antenna, while a parabolic compound lens immersed in index matching liquid in contact with the diamond collects the fluorescence into an analog photodetector. Filters are placed between the parabolic compound lens and the detector in order to remove the remaining excitation laser from the fluorescence. The magnetic field before the measurement is characterized to benchmark the performance of the protocol. The signals are generated at the analog outputs of an arbitrary waveform generator (AWG), amplified in identical MW amplifiers $\sim 43$dB (ZHL-16W-43-S+, Mini-circuits, USA) and finally connected to the antenna ports. To compensate for differences in signal paths due to different cable length, differences in amplifier behavior or different responses of the antenna ports, the amplitude and phase of the second AWG channel (Q) is swept while maintaining the first channel (I) constant and a continuous wave ESR is performed with each configuration, resulting in the results in Fig.~\ref{fig:fig2} of the main text. The port where the constant AWG channel is connected sets the direction of the $X$ axis (I channel) while the out of phase channel (Q) port sets the $Y$ axis of our laboratory frame of reference.


\subsection*{Supplementary Text}

\subsubsection*{Effect of circular MW fields on NV centers}
This section shows the explicit effect of phase difference between MW fields in the control of the electronic spin of NV centers. For this purpose, we start from the Hamiltonian described in the main text:
\begin{equation}\label{Phase modulated pulses}
     H = D S_z^2 + \gamma_e \vec{B} \cdot \vec{S}+ \Omega' \left(S_x \cos{\left( \omega t + \varphi_1\right)} + S_y \sin{\left( \omega t + \varphi_2\right)}\right) \ \ .
\end{equation}

Moving into interaction picture with respect to $ \omega S_z^2$ and defining $\Omega = \frac{\Omega'}{\sqrt{2}}$ the resulting Hamiltonian after the rotating wave approximation becomes:

\begin{equation} \label{Phase dependent pulses}
    H' = D S_z^2 + \gamma_e B_z S_z - \omega S_z^2  + \frac{\Omega}{2} \begin{pmatrix}
        0 & e^{-i\varphi_1} + e^{-i\varphi_2} & 0 \\
        e^{i\varphi_1} + e^{i\varphi_2} & 0 & e^{i\varphi_1} - e^{i\varphi_2} \\
        0 & e^{-i\varphi_1} - e^{-i\varphi_2} & 0
    \end{pmatrix} \ \ .
\end{equation}
Equation (\ref{Phase dependent pulses}) shows the effect of MW handedness on electronic transitions. When $\varphi_1 = \varphi_2$, only terms coupling $\ket{0}\longleftrightarrow \ket{+1}$ will survive in the Hamiltonian generating transitions only between those states, while decoupling state $\ket{-1}$ from the interaction. This corresponds to a right-handed circularly polarized MW field around the NV axis. On the contrary, choosing $\varphi_2  = \varphi_1 + \pi$ only the terms coupling $\ket{0}\longleftrightarrow \ket{-1}$ will survive, decoupling state $\ket{1}$, which corresponds to a left-handed MW polarization.

\subsubsection*{Peak selective MW polarizations}

The optimal microwave (MW) ellipticities to maximally attenuate the resonance lines produced by NV centers in a single crystal axis can be calculated following the procedure described in the main text. Here the explicit details and considerations of a flat and a tilted diamond in the antenna are considered. To find explicit expressions for the ellipticities, we describe the NV axis in the laboratory frame formed by taking the $X$ and $Y$ directions along the MW ports. As described in  the main text, choosing this set of coordinates the axis orientations contain mixed terms in the $XYZ$ directions. To ease calculations we choose a new set of coordinates $X_{nv}, Y_{nv}$ oriented along the diamond crystal diagonals as described in equation (\ref{Diagonal reference frame}).

\begin{equation} \label{Diagonal reference frame}
    \begin{matrix}
        X_{nv} = \frac{1}{\sqrt{2}} \left( X + Y \right) \\
        Y_{nv} = \frac{1}{\sqrt{2}} \left( Y -X \right)
    \end{matrix}
\end{equation}

\begin{figure}
  \centering
  \includegraphics[height=6cm]{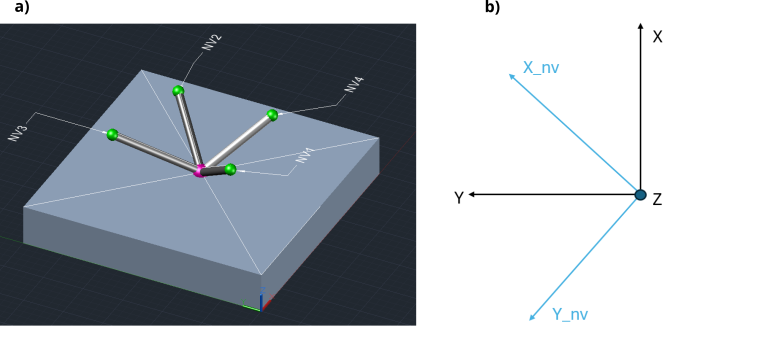}
  \caption{\textbf{Schematic representation of the crystal directions and reference frames.} $\boldsymbol{a)}$ Axis orientations in diamond. $ \boldsymbol{b)}$ Chosen coordinates for NV reference frame (blue) and laboratory frame of reference (black).}
  \label{fig: NV axis orientation}
\end{figure}

The orientations of NV directions, together with the lab frame and the newly chosen set of coordinates can be seen in figure (\ref{fig: NV axis orientation}). To differentiate between laboratory frame and the diagonal frame we denote all vectors written in the former with $nv$. The NV axis in the $X_{nv},Y_{nv}$ reference frame are:

\begin{equation}
	\begin{matrix}
        \hat{u_1}^{nv} = \left( \frac{\sqrt{2}}{\sqrt{3}}, 0, \frac{1}{\sqrt{3}}\right) \ , \\
        \hat{u_2}^{nv} = \left( -\frac{\sqrt{2}}{\sqrt{3}}, 0, \frac{1}{\sqrt{3}}\right) \ , \\
        \hat{u_3}^{nv} = \left(  0, \frac{\sqrt{2}}{\sqrt{3}}, \frac{1}{\sqrt{3}}\right) \ , \\
        \hat{u_4}^{nv} = \left(  0, - \frac{\sqrt{2}}{\sqrt{3}}, \frac{1}{\sqrt{3}}\right) \ .
    \end{matrix}
\end{equation}

Using this description, we show the procedure to calculate orthogonal MW circular polarizations for each axis. The objective is to find the MW signals in the $X$ and $Y$ direction in the lab frame so that the effect of those polarizations is to generate a circular polarization orthogonal to a single NV axis with a modulation along that axis. If this can be done for each NV axis it will be possible to individually select each of the peaks in  the ESR signal of an ensemble of NVs using only in-plane MW signals.\\

As described in the main text, the first step for this purpose is to find 2 orthonormal vectors to each NV axis. This is easiest to do using the $XY_{nv}$ coordinates. In this frame of reference the NV orientations come in pairs with components in $X_{nv}$ and $Z$ or $Y_{nv}$ and $Z$ respectively, $\hat{u_{1,2}}^{nv} = \left( n^x, 0, n^z \right) $ and $\hat{u_{3,4}}^{nv} = \left( 0, n^y, n^z \right) $ . The orthogonal vectors for each orientation can be written as follows:

\begin{equation}
    \begin{cases}
        \vec{M}_{1,2} = \left(0,1,0 \right)_{nv} \\
        \vec{N}_{1,2} = \left(-n^z, 0, n^x \right)_{nv}
    \end{cases}
    \\
    \begin{cases}
        \vec{M}_{3,4} = \left(1,0,0 \right)_{nv} \\
        \vec{N}_{3,4} = \left(0, n^z, -n^y \right)_{nv}
    \end{cases}
\end{equation}

Following equation (\ref{Phase modulated pulses}), around each axis the right and left-handed ($\vec{r}, \vec{l}$ respectively) circular polarizations are defined as in the main text:

\begin{equation}
    \begin{cases}
        \vec{r_i} = \cos(\omega t) \vec{M_i} + \sin(\omega t) \vec{N_i} \\
        \vec{l_i} = \cos(\omega t) \vec{M_i} - \sin(\omega t) \vec{N_i}
    \end{cases}
\end{equation}

However, this will not be possible to achieve with an antenna that can only generate fields in the $XY$ plane of the laboratory, since some of these vectors orthogonal to the NV axis will have components in the $Z$ direction. The proposed method to overcome this difficulty is to generate MW polarizations in the $XY$ lab frame that generate the mentioned $\vec{r}, \vec{l}$ signals over the axis together with a term parallel to the NV axis. Microwaves will only be able to rotate the spin and generate the change in fluorescence if they are orthogonal to the NV axis, while the parallel component is expected to modulate the resonance frequencies of the energy levels. This is expected to decrease the resolution of the peaks by generating line broadening, however, because of the fast oscillating microwave signals this effect is expected to average over acquisitions.\\

The vectors that need to be constructed can be written as:

\begin{equation}
    \begin{cases}
        \vec{V_i} = \left( a \vec{r_i} + b \hat{u_i} \right) \\
        \vec{T_i} = \left( a \vec{l_i} + b \hat{u_i} \right) \ \ .
    \end{cases}
\end{equation}

In this case the coefficient $a$ will be the strength of the microwave fields generating the spin rotations, and thus, proportional to the Rabi frequencies of the NV. On the other hand, $b$ will be a coefficient that needs to be adjusted so that the MW needed to generate these signals have no $Z$ component in the lab frame. The vectors obtained in the NV frame are the following:

\begin{equation}
    \begin{cases}
        \vec{V}^{nv}_{1,2} = \left( -a n^z \sin(\omega t) + b n^x, a \cos(\omega t), a n^x \sin(\omega t) + b n^z \right) \\
        \vec{T}^{nv}_{1,2} = \left( a n^z \sin(\omega t) + b n^x, a \cos(\omega t), -a n^x \sin(\omega t) + b n^z \right) \\
       \vec{V}^{nv}_{3,4} = \left(  a \cos(\omega t), a n^z \sin(\omega t) + b n_y, -a n^y \sin(\omega t) + b n^z \right) \\
        \vec{T}^{nv}_{3,4} = \left(  a \cos(\omega t), -a n^z \sin(\omega t) + b n_y, a n^y \sin(\omega t) + b n^z \right) \ \ .
    \end{cases}
\end{equation}

Transforming this into the laboratory frame of reference the following set of vector is obtained:

\begin{equation}\label{Tilted reference frame vectors}
    \begin{cases}
        \vec{V}_{1,2} = \left( \frac{b }{\sqrt{2}} n^x -\frac{a}{\sqrt{2}}( n^z \sin(\omega t)+\cos(\omega t)) , \frac{a}{\sqrt{2}}(\cos(\omega t)- n^z \sin(\omega t)) + \frac{b}{\sqrt{2}} n^x, a n^x \sin(\omega t) + b n^z \right) \\
        \vec{T}_{1,2} = \left( \frac{a}{\sqrt{2}}( n^z \sin(\omega t)-\cos(\omega t)) + \frac{b }{\sqrt{2}} n^x, \frac{a}{\sqrt{2}}( n^z \sin(\omega t)+\cos(\omega t)) + \frac{b}{\sqrt{2}} n^x, -a n^x \sin(\omega t) + b n^z \right) \\
       \vec{V}_{3,4} = \left(  \frac{a}{\sqrt{2}}( \cos(\omega t)-n^z \sin(\omega t)) - \frac{b }{\sqrt{2}} n^y, \frac{a}{\sqrt{2}}( n^z \sin(\omega t)+\cos(\omega t)) + \frac{b }{\sqrt{2}} n^y,-a n^y \sin(\omega t) + b n^z \right) \\
        \vec{T}_{3,4} = \left( \frac{a}{\sqrt{2}}( \cos(\omega t)+n^z \sin(\omega t)) - \frac{b }{\sqrt{2}} n^y, \frac{a}{\sqrt{2}}( \cos(\omega t)-n^z \sin(\omega t)) +\frac{b }{\sqrt{2}} n^y, a n^y \sin(\omega t) + b n^z \right) \ \ .
    \end{cases}
\end{equation}

If the diamond is rotated around the $X$ or $Y$ axes, the previous reference frame becomes tilted. To take this into account and obtain the expressions of each polarization in the laboratory reference frame, that is where the antenna works, the previously achieved vectors need to be rotated. This rotation is achieved applying the following rotation matrices:\\

\begin{equation}
    R_x(\theta) = \begin{pmatrix}
        1 & 0 & 0 \\
        0 & \cos(\theta) & -\sin(\theta) \\
        0 & \sin(\theta) & \cos(\theta)
    \end{pmatrix}
\end{equation}

\begin{equation}
    R_y(\theta) = \begin{pmatrix}
        \cos(\theta) & 0 & \sin(\theta) \\
        0 & 1 & 0\\
        -\sin(\theta)&0 & \cos(\theta)
    \end{pmatrix} \ \ .
\end{equation}

\begin{figure}
    \centering
    \includegraphics[width=0.7\textwidth]{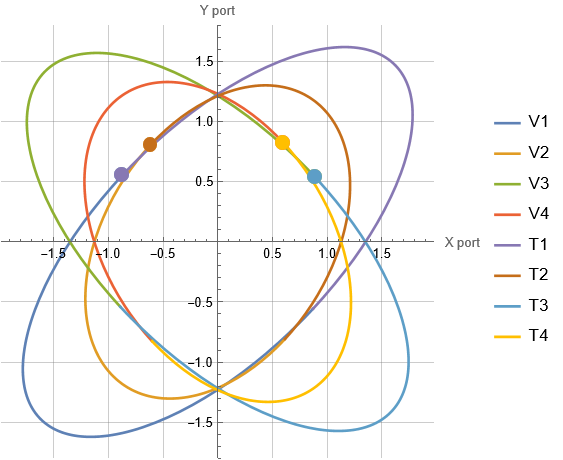}
    \caption{\textbf{Theoretical optimal spin-selective MW control fields.} MW field ellipses generating optimal circularly polarized control pulses over each NV axis for a diamond tilted by $\theta \sim -10^\circ$ and $\phi \sim 1^\circ$ calculated using the expressions from Eqs.~(\ref{signal port X formula Main}) and (\ref{signal port Y formula Main}) of the main text. The points indicate the starting points of the ellipses and half of their period is drawn to identify right- and left-handed polarizations.}
    \label{fig: Tilted diamond MW polarizations}
\end{figure}

This allows to calculate a set of microwave polarizations that will select exactly each of the peaks created by each axis direction. When the diamond is tilted the vectors calculated before become tilted, therefore any vector in the tilted diamond frame from now on is denoted with a $t$. To calculate the shape of each vector in the laboratory frame the rotation matrix needs to be applied $\vec{U}_{lab} = R_x(\phi )R_y(\theta) \vec{U}^{t}$, where $\vec{U}^{t}$ can be any of the $\vec{V}_i, \vec{T}_{i}$ in equation (\ref{Tilted reference frame vectors}) that now become $\vec{V}_i^t, \vec{T}_{i}^t$ since they are tilted. Because the antenna only creates fields in the $X$ or $Y$ direction of the lab frame the $Z$ component of the vectors $\vec{U}_{lab}$ must be canceled. This is where the component along the axis direction of magnitude $b$ comes into play. As mentioned above, this component will not rotate the spin, but instead make a small modulation of the resonance frequency. Therefore, its magnitude can be chosen freely without affecting the effectiveness of the MW signals in selecting resonant peaks. The magnitude $b$ that satisfies $\vec{U}_{lab} \cdot \hat{z} = 0$ needs to be found. After solving for $b$ and plugging the solved values into the components of the vectors in the laboratory reference frame, the expressions for the signals in each port for a tilted diamond from main text are obtained. The corresponding coefficients for Eqs.~(\ref{signal port X formula Main}) and (\ref{signal port Y formula Main}) in the main text are:

\begin{table}[h!]
\centering
\begin{tabular}{|c|c|c|c|c|}
\hline
 &\textbf{$V_{1,2}$} & \textbf{$T_{1,2}$} & \textbf{$V_{3,4}$} & \textbf{$T_{3,4}$} \\ 
\hline
$s_i$ & -1 & +1 & -1 & +1 \\ 
\hline
$p_i$ & -1 & +1 & +1 & +1 \\ 
\hline
$q_i$ & -1 & -1 & +1 & +1 \\ 
\hline
$g_i$ & +1 & -1 & +1 & +1 \\ 
\hline
$h_i$ & +1 & +1 & +1 & -1 \\ 
\hline
$f_i$ & -1 & -1 & +1 & +1 \\ 
\hline
\end{tabular}
\caption{Values of the coefficients in Eqs.~(\ref{signal port X formula Main}) and (\ref{signal port Y formula Main}) of the main text.}
\label{tab:my_table}
\end{table}
The difference of each individual signal and their polarization is shown in figure (\ref{fig: Tilted diamond MW polarizations}). The exact MW phase configuration to attenuate each transition in the experimental setup where $\theta \sim -10 ^\circ, \phi \sim 1^\circ$ are given in Table \ref{tab:Phase values theory}.\\
\\

\begin{table}[h!]
\centering
\begin{tabular}{|c|c|c|}
\hline
 \text{Ellipse} & \text{Phase difference (${}^{\circ}$)} & \text{Q/I gain} \\
 \hline
 \text{NV1 (+)} & 311 & 0.91 \\
 \hline
 \text{NV2 (+)} & 291 & 1.07 \\
 \hline
 \text{NV3 (+)} & 231 & 0.90 \\
 \hline
 \text{NV4 (+)} & 248 & 1.09 \\
 \hline
 \text{NV1 (-)} & 49 & 0.91 \\
 \hline
 \text{NV2 (-)} & 69 & 1.07 \\
 \hline
 \text{NV3 (-)} & 129 & 0.90 \\
 \hline
 \text{NV4 (-)} & 112 & 1.09 \\
 \hline
\end{tabular}
\caption{Values of the phase between the signals in ports $X$ and $Y$ from Eqs.~(\ref{signal port X formula Main}) and (\ref{signal port Y formula Main}) for the optimal polarizations of a tilted diamond.}
\label{tab:Phase values theory}
\end{table}

These calculations are consistent with the experimental results shown in the main text. Here we present the optimal MW configurations targeting all peaks from the ESR, instead of the only $4$ necessary. Note that the intrinsic tilt of the diamond in the setup together with the strict phase requirements of optimal attenuation allow to find individual MW configurations for each axis.

\subsubsection*{Flat diamond ($\theta, \phi  = 0 ^\circ $) signals and handedness relations}

When the tilt angles are null $\theta, \phi =  0$, the polarizations affecting NVs $1,2$ and $3,4$ respectively will become degenerate. This simplified particular case makes for a good starting point to study the different signals. When the diamond is placed flat over the antenna, the expressions in Eqs.~(\ref{signal port X formula Main}) and (\ref{signal port Y formula Main}) from the main text simplify due to the cancellation of all terms with $\sin(\theta)$ or $\sin(\phi)$. The new vectors become:

\begin{equation} 
	V_{1,2} = \begin{pmatrix}
                -\frac{a}{\sqrt{2}} \left( \frac{\sin(\omega t)}{n_z} +  \cos(\omega t) \right) \\
                
                -\frac{a}{\sqrt{2}}  \left( \frac{\sin(\omega t)}{n_z} + \cos(\omega t) \right) \\

                0
                \end{pmatrix}
\end{equation} \ \break
\begin{equation}
    V_{3,4} = \begin{pmatrix}
                \frac{a}{\sqrt{2}}  \left(\cos(\omega t)- \frac{\sin(\omega t)}{n_z}   \right) \\
                
                \frac{a}{\sqrt{2}} \left(\cos(\omega t)+\frac{\sin(\omega t)}{n_z}   \right) \\

                0
    \end{pmatrix}
\end{equation}

\begin{equation} \label{R pol NV1,2 flat}
    T_{1,2} = \begin{pmatrix}
                \frac{a}{\sqrt{2}} \left( \frac{\sin(\omega t)}{n_z} - \cos(\omega t) \right) \\
                
                \frac{a}{\sqrt{2}}  \left( \frac{\sin(\omega t)}{n_z} + \cos(\omega t) \right) \\

                0
    \end{pmatrix}
\end{equation} \ \break
\begin{equation}\label{L pol NV3,4 flat}
    T_{3,4} = \begin{pmatrix}
                \frac{a}{\sqrt{2}}  \left(\cos(\omega t)+ \frac{\sin(\omega t)}{n_z}   \right) \\
                
                \frac{a}{\sqrt{2}} \left(\cos(\omega t)-\frac{\sin(\omega t)}{n_z}   \right) \\

                0
    \end{pmatrix} \ \ .
\end{equation}

These can be rewritten as sine waves with a fixed phase difference. Taking the signal in the $X$ direction as the reference the previous vectors can be written as follows:

\begin{equation} \label{R pol NV1,2 flat phase}
    V_{1,2} = \begin{pmatrix}
                \sqrt{2} a \sin(\omega t) \\
                
                \sqrt{2} a \sin(\omega t-\frac{\pi}{3})\\

                0
    \end{pmatrix}
\end{equation} \ \break
\begin{equation}
    V_{3,4} = \begin{pmatrix}
                \sqrt{2} a \sin(\omega t) \\
                
                \sqrt{2} a \sin(\omega t+\frac{4\pi}{3})\\

                0
    \end{pmatrix}
\end{equation}

\begin{equation} 
    T_{1,2} = \begin{pmatrix}
                \sqrt{2} a \sin(\omega t) \\
                
                \sqrt{2} a \sin(\omega t+\frac{\pi}{3})\\

                0
    \end{pmatrix}
\end{equation} \ \break
\begin{equation} \label{L pol NV3,4 flat phase}
    T_{3,4} = \begin{pmatrix}
                \sqrt{2} a \sin(\omega t) \\
                
                \sqrt{2} a \sin(\omega t+\frac{2\pi}{3})\\

                0
    \end{pmatrix} \ \ .
\end{equation}

\begin{figure}
    \centering
    \includegraphics[width=\textwidth]{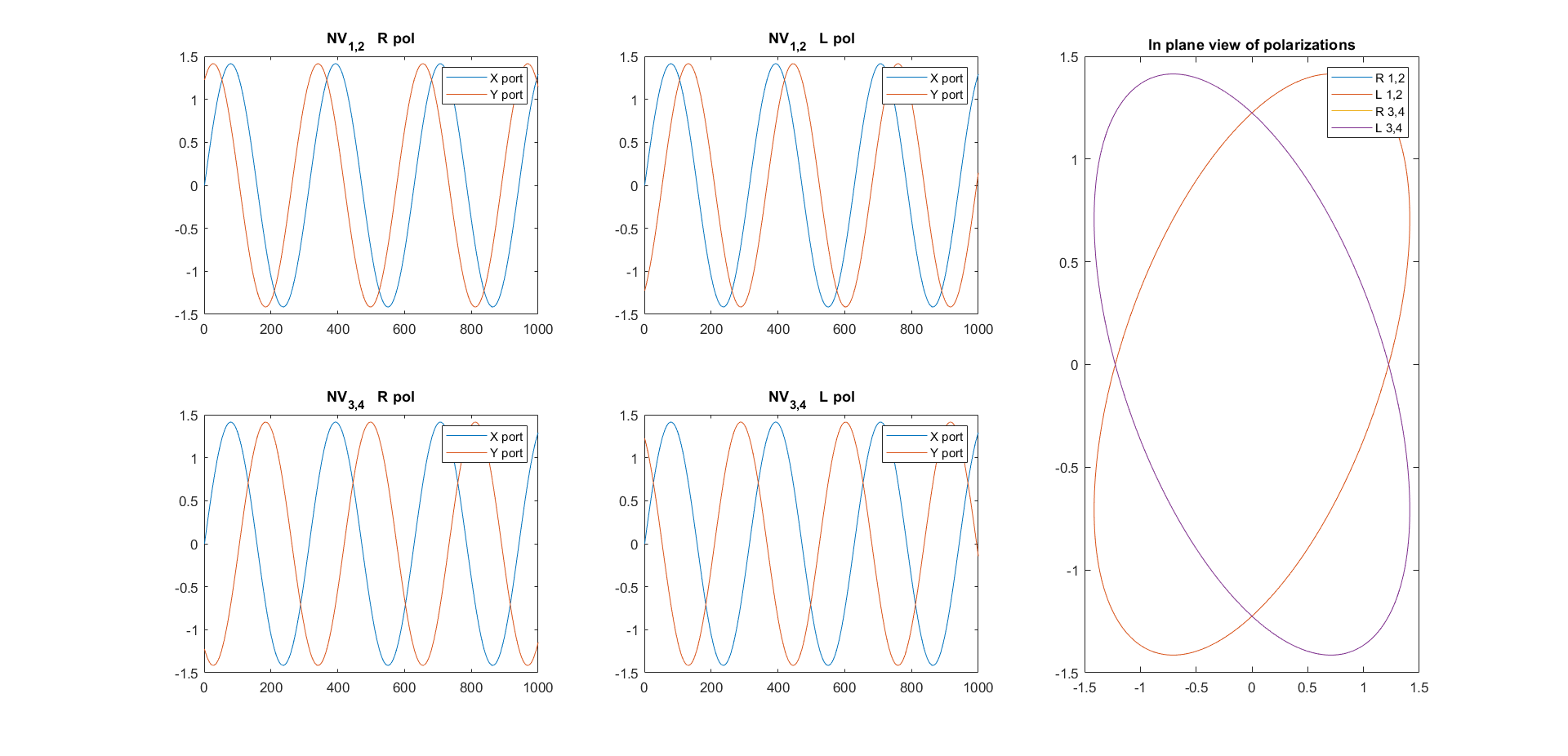}
    \caption{\textbf{Optimal MW fields for flat diamond sample.} Signals to select transitions peaks in ESR spectrum for the different NV orientations. Because the diamond is not tilted now the symmetry of the system makes that the peaks can only be selected by pairs. Notice the right-handed polarizations in the in-plane view are the same as the left handed, but have been covered by these when plotting. }
    \label{fig:MW signals for flat diamond}
\end{figure}

These expressions show the relation in phase and amplitude needed between the signal in each port of the antenna to generate the peak and axis-selective polarizations. In this non-tilted diamond scenario to target the peak corresponding to the transition $m_s =0,1$ NVs 1 and 2 require that the signal in the $Y$ direction has a $-\frac{\pi}{3}$ phase and the same amplitude, while for NVs $3,4$ the dephase must be of $\frac{4\pi}{3}$. To target the transitions $\ket{0} \longrightarrow \ket{-1}$ NVs $1,2$ require the same amplitude in both directions but the $Y$ direction must be out of phase by $\frac{ \pi}{3}$, while for NVs $3,4$ this direction must be out of phase by $\frac{2 \pi}{3}$ while maintaining the same amplitude. Notice, the power that will be inverted in rotating the spin will be a fraction $\frac{1}{\sqrt{2}}$ of the amplitude introduced in each port. These signals are visualized in Fig. \ref{fig:MW signals for flat diamond}.

\subsubsection*{Diamond flat on antenna experimental verification}
In order to test the effect of diamond tilt with respect to the antenna and find complete peak attenuation, the calibration steps described in the main text were repeated using a home-made confocal microscope.

\begin{figure}
    \centering
    \includegraphics[width=\textwidth]{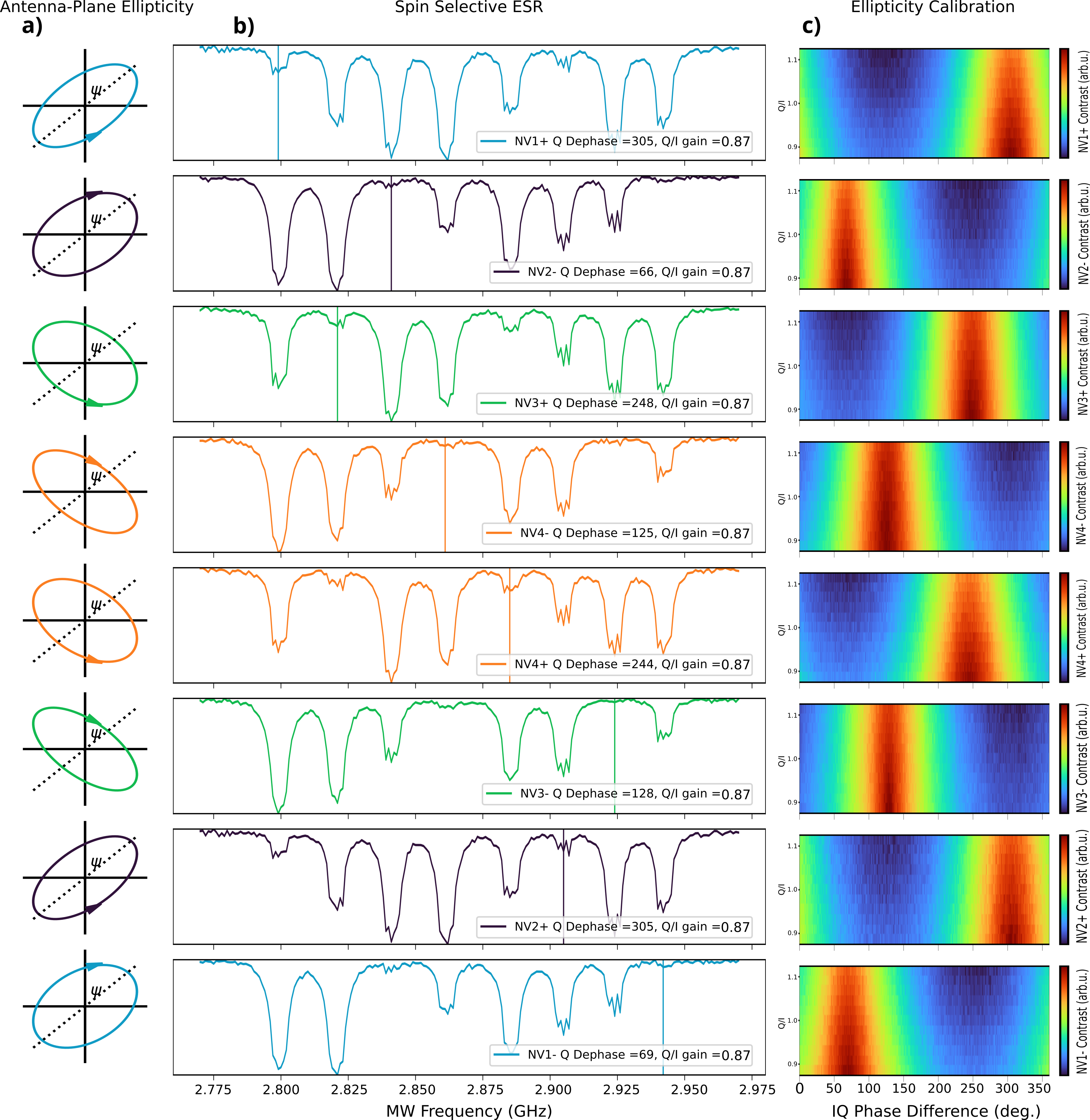}
    \caption{\textbf{Results of the calibration using a confocal microscope and a flat diamond over the antenna}. \textbf{a)} optimal spin-selective elliptic MW fields in the antenna plane. \textbf{b)} Record of the eight optimal MW field configurations to attenuate each peak in the ESR measurement. Because the diamond is almost flat on the antenna, peaks are attenuated by pairs. \textbf{c)} Map of the phase difference and intensity fractions on the Q, I ports of the MW antenna, showing the appearance of the optimal attenuation of each transition line.}
    \label{fig:Confocal callibration}
\end{figure}

The use of the confocal microscope allowed to obtain a more precise placement of the diamond in the center of the antenna, thanks, to the higher accessibility to the sample. Besides, the small confocal volume used for excitation and fluorescence collection means the MW fields in the interrogated area are substantially more homogeneous than in the experiments shown in the main text. The results of the calibration are shown in figure (\ref{fig:Confocal callibration}). From the found optimal MW configurations the effect of the tilt can be observed. Compared to the measurement in the main text, where the diamond was tilted, peaks are attenuated by pairs. The MW configuration attenuating one peak always has almost the same effect on another peak as expected from the theoretical results for a flat diamond. This confirms the need of some tilt in the diamond to uniquely resolve the vector magnetic field with independent phase configurations. Besides, the smaller collection and excitation volume resulting in more homogeneous signals gets reflected in the optimal peak attenuation. Most peaks in these measurements get almost perfectly attenuated, meaning that all NV centers in each axis are affected by the same MW fields which can be set to be completely circular. Looking at the optimal phases the slight differences in phase ($\sim 5 ^\circ$) can be explained by a small tilt in the diamond due to imprecise placing in the antenna, and errors in the definition of the optimal MW field due to noise in the measurement.

\subsubsection*{Phase dependence of peak attenuation}

\begin{figure}
    \centering
    \includegraphics[width=0.9\linewidth]{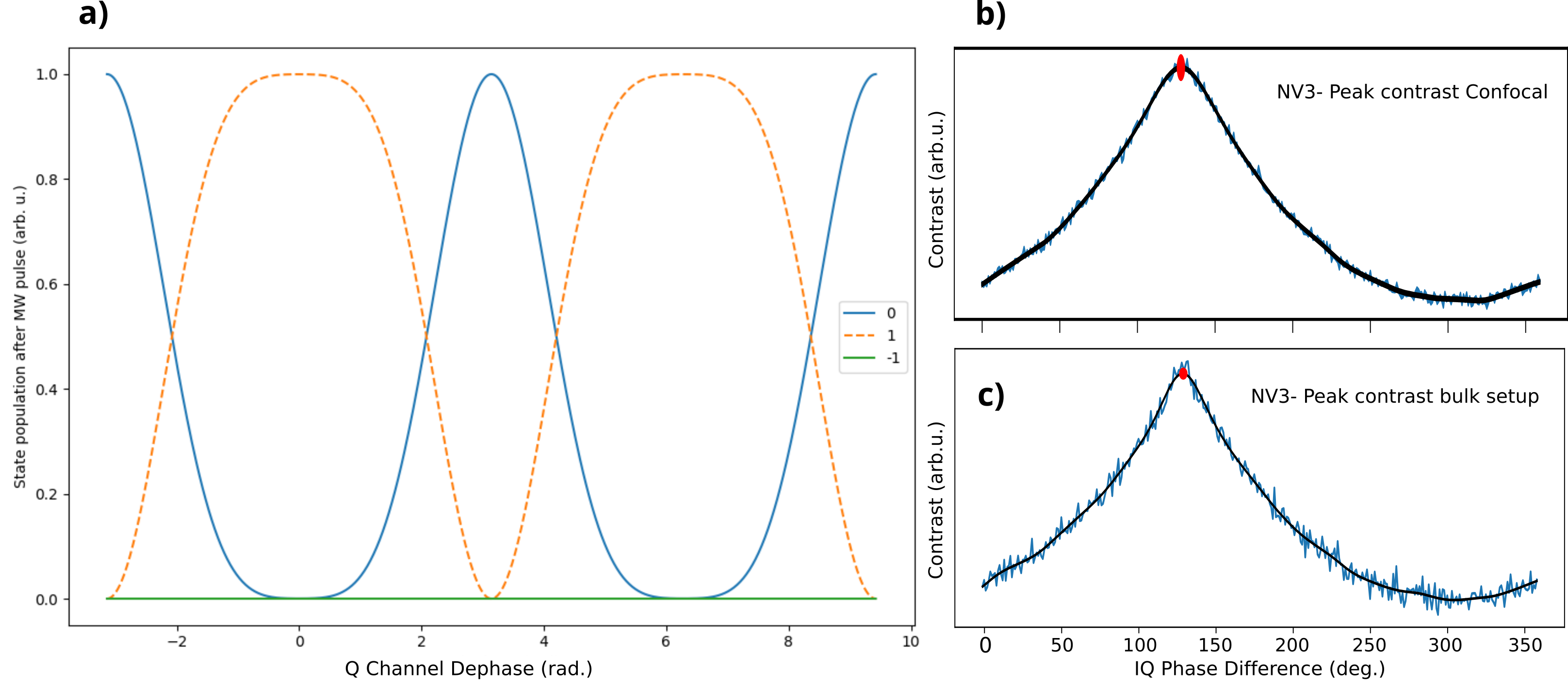}
    \caption{\textbf{Theoretical and experimental phase dependence of peak attenuation}. \textbf{a)} Simulation of the effect of phase in the population inversion attenuation of a $\pi$-pulse. \textbf{b)} (\textbf{c)}) Dependence of the peak attenuation on MW phase for the flat diamond confocal microscope (bulk tilted diamond) measurements. The black line represents the filtered signal using a Butterworth filter to find the optimal phase configuration over the signal noise.}
    \label{fig:phase dependency simulation and meas}
\end{figure}

To verify the sharp dependence of peak attenuation with the phase between MW ports, we perform simulations of a MW pulse over NV centers. Starting from the Hamiltonian in equation (\ref{Phase dependent pulses}) we set a fixed value of $\Omega$ and evolve state $\ket{0}$ under the time evolution operator of a $\pi$-pulse $U_\pi = \exp{\left(- i  H '(\varphi) \frac{1}{2 \Omega} \right)}$ choosing $\omega = D + \gamma_e B$. The population of states $\ket{0}, \ \ket{\pm 1}$ are readout at the end of the pulse for different values of the phase difference between MW ports ($\varphi$). The results are shown in Fig. \ref{fig:phase dependency simulation and meas}. The simulation shows the probability of a transition occurring during the $\pi$-pulse is high for a wide range of phase difference values, while in a narrow region close to perfect circular polarization the transition is prohibited resulting in peak attenuation in an ESR measurement. The theoretical results reflect the experimental results in this work, where peak attenuation is sharply dependendent on MW phase difference (Fig. \ref{fig:phase dependency simulation and meas}).



\end{document}